\documentclass[letterpaper]{article}
\usepackage{aaai/aaai}
\usepackage{times}
\usepackage{helvet}
\usepackage{courier}

\usepackage{graphicx}
\usepackage{amsmath}
\usepackage{tikz}

\usepackage{tabu}
\usepackage{array}

\usepackage[bottom]{footmisc}

\usepackage[inline]{enumitem}

\def\etal{et~al.\,}
\def\ie{\emph{i.e.}\,}
\def\eg{\emph{e.g.}\,}

\newif\ifappendix
\newif\ifnoappendix 
\noappendixtrue 

\makeatletter
\newcommand\footnoteref[1]{\protected@xdef\@thefnmark{\ref{#1}}\@footnotemark}
\makeatother

\begin{document}

 
\title{Adaptive Representations for Tracking Breaking News on Twitter}

\author{Igor Brigadir \and Derek Greene \and P\'{a}draig Cunningham\\
Insight Centre for Data Analytics\\
University College Dublin\\
igor.brigadir@ucdconnect.ie, derek.greene@ucd.ie, padraig.cunningham@ucd.ie\\
}

\maketitle

\begin{abstract}
\begin{quote}
\noindent 
Twitter is often the most up-to-date source for finding and tracking breaking news stories. Therefore, there is considerable interest in developing filters for tweet streams in order to track and summarize stories. This is a non-trivial text analytics task as tweets are short, and standard retrieval methods often fail as stories evolve over time. In this paper we examine the effectiveness of adaptive mechanisms for tracking and summarizing breaking news stories. We evaluate the effectiveness of these mechanisms on a number of recent news events for which manually curated timelines are available. Assessments based on ROUGE metrics indicate that an adaptive approaches are best suited for tracking evolving stories on Twitter.
\end{quote}
\end{abstract}

\section{Introduction} 
Manually constructing timelines of events is a time consuming task that requires considerable human effort. Twitter has been shown to be a reliable platform for breaking news coverage, and is widely used by established news wire services. While it can provide an invaluable source of user generated content and eyewitness accounts, the terse and unstructured language style of tweets often means that traditional information retrieval have difficulty with this type of content.

Recently, Twitter has introduced the ability to construct \emph{custom timelines}\footnote{blog.twitter.com/2013/introducing-custom-timelines} or \emph{collections} from arbitrary tweets. The intended use case for this feature is the ability to curate relevant and noteworthy tweets about an event or topic.

We propose an approach for constructing \emph{custom timelines} incorporating distributional semantic language models (DSMs) trained on tweet text. DSMs create useful representations of terms used in tweets, capturing the syntactic and semantic relationships between words. 

We evaluate several retrieval approaches including a neural network language model introduced by Mikolov~\etal \shortcite{mikolov2013efficient}, Random Indexing \cite{randomindexing}, and a BM25 based method.
 
Usually, DSMs are trained on large static data sets. In contrast, our approach trains models on relatively smaller sets, updated at frequent intervals. Regularly retraining using recent tweets allows our proposed approach to adapt to temporal drifts in content.

This retraining strategy allows us to track a news event as it evolves, since the vocabulary used to describe it will naturally change as it develops over time. Given a seed query, our approach can automatically generate chronological timelines of events from a stream of tweets, while continuously learning new representations of relevant words and entities as the story changes. Evaluations performed in relation to a set of real-world news events indicate that adaptive approaches allow us to track events more accurately, when compared to nonadaptive models (models that rely on large, static data sets).

\section{Problem Formulation} 
%
\emph{Custom timelines}, curated tweet collections on \emph{Storify}\footnote{www.storify.com}, and liveblog platforms such as \emph{Scribblelive}\footnote{www.scribblelive.com} are conceptually similar and are popular with many major news outlets.

For the most part, liveblogs and timelines of events are manually constructed by journalists. Rather than automating construction of timelines entirely, our proposed approach offers editorial support for this task, allowing smaller news teams with limited budgets to use resources more effectively. Our contribution focuses on retrieval and tracking rather than new event detection or verification.

We define a timeline of an event as a timestamped set of tweets relevant to a query, presented in chronological order. The problem of adaptively generating timelines for breaking news events is cast as a topic tracking problem, comprising of two tasks:

\subsubsection{Realtime ad-hoc retrieval:} For each target query (some keywords of interest), retrieve all relevant tweets from a stream posted after the query. Retrieval should maximize recall for all topics (retrieving as many possibly relevant tweets as available).

\subsubsection{Timeline Summarization:} Given all retrieved tweets relating to a topic, construct a timeline of an event that includes all detected aspects of a story. Summarization involves removal of redundant or duplicate information while maintaining good coverage.

\section{Related Work}
The problem of generating news event timelines is related to topic detection and tracking, and multi-document summarization, where probabilistic topic modelling approaches are popular. Our contribution attempts to utilise a state-of-the-art neural network language model (NNLM) and other distributional semantic approaches in order to capitalise on the vast amount of microblog data, where semantic concepts between words and phrases can be captured by learning new representations in an unsupervised manner.
\subsubsection{Timeline Generation.}
An approach by  Wang \shortcite{Wang} that deals with longer news articles, employed a Time-Dependent Hierarchical Dirichlet Model (HDM) for generating timelines using topics mined from HDM for sentence selection, optimising coverage, relevance, and coherence. Yan~\etal \shortcite{Yan} proposed a similar approach, framing the problem of timeline generation as an optimisation problem solved with an iterative substitution approach, optimising for diversity as well as coherence, coverage, and relevance. Generating timelines using tweets was explored by Li \& Cardie \shortcite{Li}. However, the authors solely focused on generating timelines of events that are of a personal interest.
\emph{Sumblr} \cite{Shou2013} uses an online tweet stream clustering algorithm, which can produce summaries over arbitrary time durations, by maintaining snapshots of tweet clusters at differing levels of granularity.
\subsubsection{Tracking News Stories.}
To examine the propagation of variations of phrases in news articles, Leskovec~\etal \shortcite{Leskovec2009} developed a framework to identify and adaptively track the evolution of unique phrases using a graph based approach.
In \cite{Chong2009}, a search and summarization framework was proposed to construct summaries of events of interest. A Decay Topic Model (DTM) that exploits temporal correlations between tweets was used to generate summaries covering different aspects of events.
Osborne \& Lavrenko \shortcite{Osborne2002} showed that incorporating paraphrases can lead to a marked improvement on retrieval accuracy in the task of First Story Detection.
\subsubsection{Semantic Representations.}
There are several popular ways of representing individual words or documents in a semantic space. Most do not address the temporal nature of documents but a  notable method that does is described by Jurgens and Stevens \shortcite{Jurgens2009}, adding a temporal dimention to Random Indexing for the purpose of event detection. Our approach focuses on summarization rather then event detection, however the concept of using word co-occurance to learn word representations is similar. 

\section{Source Data}
The corpus of tweets used in our experiments consists of a stream originating from a set of manually curated ``newsworthy'' accounts created by journalists\footnote{Tweet data provided by \emph{Storyful} (www.storyful.com)} as Twitter lists. Such lists are commonly used for monitoring activity and extracting eyewitness accounts around specific news stories or regions.

Our stream collects tweets from a total of 16,971 unique users, segmented into 347 geographical and topical lists. This sample of users offers a reasonable coverage of potentially newsworthy tweets, while reducing the need to filter spam and personal updates from accounts that are not focused on disseminating breaking news events. While these lists of users have natural groupings (by country, or topic), we do not segment the stream or attempt to classify events by type or topic.


\section{Event Data}
\label{sec:eventdata}
As ground truth for our experiments, we use a set of publicly available \emph{custom timelines} from Twitter, relevant content from \emph{Scribblelive} liveblogs, and collections of tweets from \emph{Storify}. Multiple reference sources are included when available.

It is not known what kind of approach was used to construct these timelines, but as our stream includes many major news outlets, we expect some overlap with our sources, although other accounts may be missing. Our task involves identifying similar content to event timelines posted during the same time periods.

Since evaluation is based on content, reference sources may contain information not in our dataset and vice versa. Where there were no quoted tweets in ground truth, the text was extracted as a sentence update instead. Photo captions and other descriptions were also included in ground truth. Advertisements and other promotional updates were removed.

For initial model selection and tuning, timelines for six events were sourced from Twitter and other live blog sources:
\begin{itemize}
  \item ``BatKid'': Make-A-Wish foundation event.
  \item ``Iran'': Follows Iranian Nuclear proliferation talks.
  \item ``LAX'': A shooting at LAX.
  \item ``RobFord'': Senator Rob Ford Council meeting.
  \item ``Tornado'': Reports of multiple tornadoes in US midwest.
  \item ``Yale'': An Alert regarding a possible gunman at Yale University.
\end{itemize}

These events were chosen to represent an array of different event types and information needs. Timelines range in length and verbosity as well as content type. See Table \ref{fig:evalevents}.

``Batkid'' can be characterised as a rapidly developing event, but without contradictory reports. ``Yale'' is also a rapidly developing event, but one where confirmed facts were slow to emerge. ``Lax'' is a media heavy event spanning just over 7 hours while ``Tornado'' spans 9 hours and is an extremely rapidly developing story, comprised mostly of photos and video of damaged property. ``Iran'' and ``Robford'' differ in update frequency but are similar in that related stories are widely discussed before the evaluation period.

In some cases the same tweets present in a human generated timeline appeared in our automatically generated timelines (see Table~\ref{fig:tweets}), providing an indication that our data source provides good coverage of newsworthy sources for a variety of events.

\begin{table}
\small
\begin{tabular}{|l|l|l|}\hline
\multicolumn{1}{|m{0.83cm}|}{Event Period} & Ground Truth & Retrieved Tweets (NNLM) \\\hline

\multicolumn{1}{|m{0.8cm}|}{15:30 to 16:11} & \multicolumn{1}{m{3.2cm}|}{\raggedright Confirmed report of a person w/ gun on/near Old Campus. SHELTER IN PLACE.} & \multicolumn{1}{m{3.4cm}|}{\raggedright NOW: Police responding to reports of a person with a gun at Yale University. Shelter in Place issued on Central Campus (via @Yale)} \\\hline

\multicolumn{1}{|m{0.8cm}|}{17:34 to 18:15} & \multicolumn{1}{m{3.2cm}|}{\raggedright New Haven police spokesman says there is no description of a suspect @Yale and ``This investigation is in its infancy'' \#NHV \#Yale} & \multicolumn{1}{m{3.4cm}|}{\raggedright New Haven police spokesman says there is no description of a suspect @Yale  and ``This investigation is in its infancy'' \#NHV \#Yale} \\\hline

\multicolumn{1}{|m{0.8cm}|}{18:57 to 19:38} & \multicolumn{1}{m{3.2cm}|}{\raggedright hartman: possibility that witnesses of long guns saw instead law enforcement officers responding to the scene \#Yale} & \multicolumn{1}{m{3.4cm}|}{\raggedright RT @NBCConnecticut: Police say witnesses who saw person with long gun at @Yale could have seen law enforcement personnel. \#Yalelockdown} \\\hline

\end{tabular}
\caption{A manual selection of retrieved tweets for ``Yale'' event highlighting key developments, and how the adaptive NNLM model can handle concept drift with high recall.}
\label{fig:tweets}
\end{table}



For evaluation, several new events are considered:
\begin{itemize}
  \item ``MH17'': Follows shooting down of Malaysian Air Flight.
  \item ``Train'': Timeline describes a train derailment.
  \item ``Westgate'': Follows the Westgate Mall Siege.
  \item ``MH370'': Details the initial reports of the missing flight.
  \item ``Crimea'' follows an eventful day during the annexation of the Crimean peninsula.
  \item ``Bitcoin'': Reporters chase the alleged creator of Bitcoin.
  \item ``Mandela'': Reactions to illness \& death.
  \item  ``P. Walker'': Reactions to car accident \& death.
  \item ``WHCD'': White House Correspondents Dinner.
  \item ``WWDC'': Follows the latest product launches from Apple - characterised by a very high number of updates and rapidly changing context.
\end{itemize}

Table \ref{fig:evalevents} gives an overview of the reference sources, durations, content types, and update frequency for each event.

\begin{table}
\small
\begin{tabu} to \columnwidth { @{\hspace{0.3em}}X[0.25,r] |@{\hspace{0.3em}}X[1.0,l] |@{\hspace{0.3em}}X[r] |@{\hspace{0.3em}}X[r] |@{\hspace{0.3em}}X[r] |@{\hspace{0.3em}}X[0.7,r] |@{\hspace{0.3em}}X[0.8,r] }
id & Event Name: & Reference Sources: & Duration: (Hrs:min) & Total Updates & Tweets & Update Freq. \\ \hline
 & BatKid	 & 2 & 5:30 & 294 & 123 & 13.36 \\ \hline
 & Iran	 & 3 & 4:15 & 197 & 190 & 11.59 \\ \hline
 & LAX	 & 5 & 7:15 & 1186 & 944 & 40.90 \\ \hline
 & RobFord	 & 4 & 6:45 & 1219 & 904 & 45.15 \\ \hline
 & Tornado	 & 5 & 9:0 & 2224 & 1617 & 61.78 \\ \hline
 & Yale	 & 1 & 7:15 & 124 & 124 & 4.28 \\ \hline
\\ \hline
1 & MH17	 & 5 & 7:30 & 554 & 487 & 18.47 \\ \hline
2  & Train	 & 2 & 10:0 & 472 & 469 & 11.80 \\ \hline
3  & Westgate	 & 3 & 18:15 & 73 & 62 & 1.00 \\ \hline
4  & MH370	 	& 1 & 7:0 & 42 & 7 & 1.50 \\ \hline
5  & Crimea	 	 & 1 & 7:0 & 34 & 34 & 1.21 \\ \hline
6  & Bitcoin	 & 2 & 4:15 & 157 & 149 & 9.24 \\ \hline
7  & Mandela	& 2 & 4:45 & 89 & 51 & 4.68 \\ \hline
8  & WHCD	 	 & 2 & 8:0 & 617 & 440 & 19.28 \\ \hline
9  & P.Walker	  & 2 & 5:45 & 152 & 106 & 6.61 \\ \hline
10 & WWDC	 	& 2 & 3:30 & 1069 & 81 & 76.36 \\ \hline

\end{tabu}  
\caption{Details for events used for parameter fitting and evaluation. Update Frequency is average number of updates every 15 minutes.}
\label{fig:evalevents}
\end{table}

\section{Methods}
\label{sec:methods}
The task of realtime ad-hoc retrieval for constructing timelines is made challenging by the continuously updating collection of documents. Traditional approaches perform poorly lacking global term statistics (IDF counts for example) or become intractable as the collection of documents grows over time. The impact of ``cheating'' by using future term statistics in a related, but notably different realtime tweet search task is discussed in \cite{futureterms}. The key difference between the TREC realtime tweet search task, and the realtime retrieval task posed here is that the TREC task involves retrieving relevant tweets \emph{before} the query time, whereas for timeline generation, the task is to retrieve tweets posted \emph{after} the query time.

\subsection{Timeline Generation}
We compare three adaptive models: BM25 (with updating IDF component), Word2Vec and two Random Indexing approaches (with updating training data), and static variants.

In each case, we initialize the process with a query. For a given event, the tweet stream is then replayed from the event's beginning to end, with the exact dates defined by tweets in the corresponding human generated timelines. Inclusion of a tweet in the timeline is controlled by a cosine similarity with a fixed similarity threshold. The stream is processed using a fixed length sliding window updated at regular intervals in order to accommodate model training time. The fixed length sliding window approach used to build models of tweet representations (TF-IDF in BM25 or DSM in Word2Vec and Random Indexing models) means that new tweets arriving from the stream are analysed with trained models that are at most \emph{refresh rate} minutes old. Parameter settings for the \emph{window length} and refresh rates are discussed in \emph{Parameter Selection} below.

\subsubsection{Pre-processing.}
A modified stopword list was used to remove Twitter specific terms (\eg ``MT'', ``via''), together with common English stopwords. URLs and media items are removed, but mentions and hashtags are preserved. For distributional semantic models, stopwords were replaced with a placeholder token, in order to preserve relative word positions. This approach showed an improvement when compared with no stopword removal, and complete removal of stopwords. While models can be trained on any language effectively, to simplify evaluation only English tweets were considered. Language filtering was performed using Twitter metadata. On average, there are 150k-200k terms in each sliding window. Updating the sliding window every 15 minutes and retraining on tweets posted in the previous 24 hours was found to provide a good balance between adaptivity and quality of resulting representations. 

\subsubsection{Nonadaptive Approaches:} The nonadaptive representation models are variants where word vectors or term frequencies are initially trained on a large number of tweets, and no further updates to the model are made as time passes.

\subsubsection{Adaptive Approaches:} The adaptive versions use a sliding window approach to continuously build new models at a fixed interval. The trade-off between recency and accuracy is controlled by altering two parameters: \emph{window length} (\ie limiting the number of tweets to learn from) and \emph{refresh rate} (\ie controlling how frequently a model is retrained). No updates are made to the seed query, only the representation of the words changes after retraining the model.

\subsubsection{Post-processing}
For all retrieval models, to optimise for diversity and reduce timeline length the same summarization step was applied to remove duplicate and near duplicate tweets. The SumBasic\cite{vanderwende2007beyond} algorithm was chosen for producing tweet summaries with high recall\cite{inouye2011comparing}. The target length for a summary is determined by the average length of the reference summaries for an event.

\section{TF-IDF Model}
BM25\cite{JonesBM25} with microblog specific settings\cite{ferguson2011clarity} are a family of scoring functions used to rank documents according to relevance to a query. For a query $Q$ comprising of terms $q_1,\dots,q_n$ the document $D$ is scored with:

$$ \text{Score}_{bm25} (D,Q) =$$
$$ \sum_{i=1}^{n} \text{IDF}(q_i) \cdot \frac{tf(q_i, D) \cdot (k_1 + 1)}{tf(q_i, D) + k_1 \cdot (1 - b + b \cdot \frac{|D|}{\text{avgdl}})} $$

Where $tf(q_i, D)$ is the term frequency of $q_i$ in $D$, $|D|$ is document length, and $avgdl$ is average document length in the collection.

Parameter choices of $k_1$ and $b$ are set to $k_1 = 1.2, b = 0.75$. The $IDF$ calculation can also be substituted for alternatives, but most implementations, including \emph{Lucene} calculate $\text{IDF}$ of term $t$ using $\text{IDF}(t) = \log \frac{N - n(t) + 0.5}{n(t) + 0.5}$, $N$ being the total number of documents in the collection and $n(t)$ the number of documents containing $t$.

The document and term frequency counts are periodically updated as new information becomes available, using the same sliding window approach for generating training data for other models.

\section{Skip-Gram Language Model}
Recent work by \cite{Mikolov} introduced an efficient way of training a Neural Network Language Model (NNLM) on large volumes of text using stochastic gradient descent. This language model represents words as dense vectors of real values. Unique properties of these representations of words make this approach a good fit for our problem. 

The high number of duplicate and near-duplicate tweets in the stream benefits training by providing additional training examples. 
For example: the vector for the term ``LAX'' is most similar to vectors representing ``\#LAX'', ``airport'', and ``tsa agent'' - either syntactically or semantically related terms. Moreover, retraining the model on new tweets create entirely new representations that reflect the most recent view of the world. In our case, it is extremely useful to have representations of terms where ``\#irantalks'' and ``nuclear talks'' are highly similar at a time when there are many reports of nuclear proliferation agreements with Iran.%

Additive compositionality is another useful property of the these vectors. It is possible to combine several words via an element-wise sum of several vectors. There are limits to this, in that summation of multiple words will produce an increasingly noisy result. Combined with standard stopword removal, and URL filtering, and removal of rare terms, each tweet can be reduced to a few representative words. The NNLM vocabulary also treats mentions and hashtags as words, requiring no further processing or query expansion. Combining these words allows us to compare similarities between whole tweets.
%




\subsection{Training:}

The computational complexity of the skip-gram model is dependent on the number of training epochs \(E\), total number of words in the training set \(T\), maximum number of nearby words \(C\), dimensionality of vectors \(D\) and the vocabulary size \(V\), and is proportional to:
\[O=E\times T\times C\times(D+D\times\log_2(V))\]
The training objective of the skip-gram model, revisited in \cite{mikolov2013efficient}, is to learn word representations that are optimised for predicting nearby words. Formally, given a sequence of words \(w_1,w_2,\dotsc w_T\) the objective is to maximize the average log probability:
\[\frac{1}{T}\sum_{t=1}^{T} \sum_{-c\leq j\leq c, j\neq 0}^{} \log p(w_{t+j}|w_t)\]
In effect, word context plays an important part in training the model. 

\subsubsection{Pre Processing:}
For a term to be included in the training set, it must occur at least twice in the set. These words are removed before training the model.

Filtering stopwords entirely had a negative impact on overall accuracy. Alternatively, we filter stopwords while maintaining relative word positions.

Extracting potential phrases before training the model, as described in \cite{Mikolov} did not improve overall accuracy. In this pre-processing step, frequently occurring bigrams are concatenated into single terms, so that phrases like ``trade agreement'' become a single term when training a model.

\subsubsection{Training Objective:}
An alternative to the skip-gram model, the continuous bag of words (CBOW) approach was considered. The skip-gram model learns to predict words within a certain range (the context window) before and after a given word. In contrast, CBOW predicts a given word given a range of words before and after. While CBOW can train faster, skip-gram performs better on semantic tasks. Given that our training sets are relatively small, CBOW did not offer any advantage in terms of improving training time. Negative sampling from \cite{Mikolov} was not used. The context window size was set to 5. During training however, this window size is dynamic. For each word, a context window size is sampled uniformly from 1,...k. As tweets are relatively short, larger context sizes did not improve retrieval accuracy.

\subsection{Vector Representations:}

The model produces continuous distributed representations of words, in the form of dense, real valued vectors. These vectors can be efficiently added, subtracted, or compared with a cosine similarity metric.

The vector representations do not represent any intuitive quantity like word co-occurance counts or topics. Their magnitude though, is related to word frequency. The vectors can be thought of as representing the distribution of the contexts in which a word appears.

Typically, these models are trained on large, static data sets. In this case smaller sets are used, with lowered thresholds for rate terms (minimum count of 2), more training epochs and a lower learning rate. These parameters produced better performance on smaller data sets in this retrieval task, but may not be optimal for other tasks.

Vector size is also a tunable parameter. While larger vector sizes can help build more accurate models in some cases, in our retrieval task, vectors larger than 200 did not show a significant improvement in scores. \ifnoappendix (See Figure \ref{fig:vecs}) \fi

\section{Random Indexing}
Random Indexing (RI)\cite{sahlgren2005introduction} is based on seminal work on sparse distributed memory\cite{kanerva1988sparse}. While not as popular as NNLM models, it is very well suited to distributed computation, can be highly efficient and has comparable performance to more advanced techniques such as Latent Semantic Analysis/Indexing (LSA/LSI)\cite{deerwester1990indexing}.

The general approach to creating a word space model involves creating a matrix $F$ where each row $F_w$ represents a word $w$ and each column $F_c$ represents a context $c$. The context can be another co-occurring word, or a document. $F$ is then either a word-by-word or word-by-document matrix, as in the case of LSA.

These types of word spaces suffer from efficiency and scalability problems. The number of words (vocabulary) and documents can make the matrix extremely large and difficult to use. The matrix $F$ is also extremely sparse. LSA solves this dimensionality and sparsity problem with Singular Value Decomposition (SVD) though this creates other problems, as the SVD operation still requires the full co-occurrence matrix and is difficult to update with new information.

The Random Indexing approach to this problem is to use a random projection of the full co-occurrence matrix in a much lower dimensional space. Random indexing can then be thought of as a dimensionality reduction technique.

The standard Random indexing technique is a two step process: An \emph{index vector} is created for each document (or word). This index vector is still high-dimensional, though in the region of several thousands, which is still much lower than an entry in a full co-occurrence matrix. The vectors are also sparse (most entries are 0) and \emph{ternary} where randomly distributed $+1$ and $-1$ values ensure  near-orthogonality. The near-orthogonality property is an important attribute of the index vectors\cite{orthoIJIAS} in the word space created by RI.

The second step involves an element wise sum of index vectors for each co-occurrence of a word in the text. Words are then represented as d-dimensional vectors consisting of the sum of the contexts in which a word is found. Simple co-occurrence only considers immediate surrounding words, though in practice a context window is extended to include several words.

The accumulation stage results in a d-dimensional space $F_{w\times d}$ which is an approximation of the full $F_{w\times c}$ matrix. This insight is based on the Johnson-Lindenstrauss lemma\cite{frankl1988johnson} which states that distances between points are approximately preserved when projecting points into a randomly selected sub space of high dimensionality. The matrix $F$ can then be approximated by projecting (multiplying) it with a random matrix $R$:
$$F_{w\times d} R_{d\times k} = F'_{w\times k}$$

The Random Indexing approach is incremental, easily applicable to parallel computation, and efficient, involving simple integer operations.

Variants of Random Indexing approaches involve different strategies for adding index or \emph{elemental} vectors. Index vectors can also be initialized for terms rather than documents, in both cases this ``training'' step produces vectors that encode meaningful relationships between words that do not co-occur. A variant of the standard RI approach is \emph{Reflective Random Indexing} (RRI)\cite{cohen2010reflective} where vectors are built in a slightly different way.

Two variants of RI approaches are implemented as alternatives to the NNLM:

\subsubsection{Term-term RI (TTRI)}

    \begin{enumerate}[itemjoin={\newline}]
    
    	\item Assign index vector for each term.
		\item For each term, sum the index vectors for each co-occurring term in a context window.
    \end{enumerate}
    
\subsubsection{Term based Reflective RI (TRRI)}

    \begin{enumerate}[itemjoin={\newline}]
    	\item Assign index vector for each term.
    	\item Generate document vectors by summing index vectors of all terms contained in the document.
    	\item For each term, sum document vectors for each document the term occurs in.
    \end{enumerate}

The trained model represents a word space similar to the model created by the skip-gram (word2vec) model. The same additive composition approach is used to create a vector representing a whole tweet, with an element wise sum of the individual word vectors.

\section{Parameter Selection}
\label{sec:params}
Our system has a number of tuneable parameters that suit different types of events. When generating timelines of events retrospectively, these parameters can be adapted to improve accuracy. For generating timelines in real-time, parameters are not adapted to individual event types.

For all models, the \textit{seed query} (either manually entered, or derived from a tweet) plays the most significant part. Overall, for the NNLM and RI models, short event specific queries with few terms perform better than longer, expanded queries which benefit term frequency (BM25) model. In our evaluation, the same queries were used while modifying other parameters. Queries were adapted from the first tweet included in an event timeline to simulate a lack of information at the beginning of an event.

The \textit{refresh rate} parameter controls how old the training set of tweets can be for a given model. In the case of BM25 model, this affects the IDF calculations, and for NNLM and RI models, the window contains the preprocessed text used for training. As such, when the system is replaying the stream of tweets for a given event, the model used for similarity calculations is \textit{refresh rate} minutes old.

\begin{figure}

\definecolor{cd3d3d3}{RGB}{211,211,211}
\definecolor{cd5d595}{RGB}{213,213,149}
\definecolor{cffffb3}{RGB}{255,255,179}
\definecolor{caac4a4}{RGB}{170,196,164}
\definecolor{cccebc5}{RGB}{204,235,197}
\definecolor{c75b0a6}{RGB}{117,176,166}
\definecolor{c8dd3c7}{RGB}{141,211,199}
\definecolor{c6b94b0}{RGB}{107,148,176}
\definecolor{c80b1d3}{RGB}{128,177,211}
\definecolor{c9e9bb6}{RGB}{158,155,182}
\definecolor{cbebada}{RGB}{190,186,218}
\definecolor{cd2abbf}{RGB}{210,171,191}
\definecolor{cfccde5}{RGB}{252,205,229}

\begin{tikzpicture}[x=0.65pt, y=0.6pt, yscale=-1]
\input{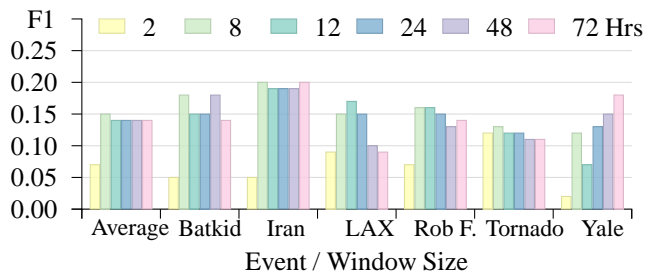}
\end{tikzpicture}
\caption{F1 scores for Adaptive model accuracy in response to changing window size}
\label{fig:windows}
\end{figure}

\textit{Window length} effectively controls how many terms are considered in each model for training or IDF calculations. While simpler to implement, this fixed window approach does not account for the number of tweets in a window, only the time range is considered. The volume of tweets is not constant over time - leading to training sets of varying sizes. However, since the refresh rate is much shorter than the window length, the natural increase and decrease in tweet volume is smoothed out. On average, there are 150k-200k unique terms in each 24 hour window. Figure~\ref{fig:windows} shows how varying window size can improve or degrade retrieval performance of different events.

Updating the sliding window every 15 minutes and retraining on tweets posted in the previous 24 hours was found to provide a good balance between adaptivity and quality of resulting representations. Larger window sizes encompassing more tweets were less sensitive to rapidly developing stories, while smaller window sizes produced noisier timelines for most events. 

\begin{figure}
\definecolor{cd3d3d3}{RGB}{211,211,211}
\definecolor{cb5b5b5}{RGB}{181,181,181}
\definecolor{cd9d9d9}{RGB}{217,217,217}
\definecolor{cd2abbf}{RGB}{210,171,191}
\definecolor{cfccde5}{RGB}{252,205,229}
\definecolor{c9e9bb6}{RGB}{158,155,182}
\definecolor{cbebada}{RGB}{190,186,218}
\definecolor{c75b0a6}{RGB}{117,176,166}
\definecolor{c8dd3c7}{RGB}{141,211,199}
\definecolor{c6b94b0}{RGB}{107,148,176}
\definecolor{c80b1d3}{RGB}{128,177,211}
\begin{tikzpicture}[x=0.65pt, y=0.6pt, yscale=-1]
\input{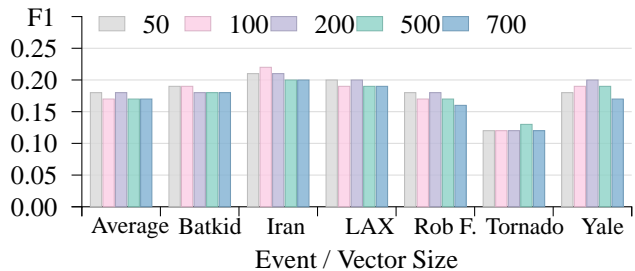}
\end{tikzpicture}
\caption{F1 scores for Adaptive model accuracy in response to changing vector size}
\label{fig:vecs}
\end{figure}

Figures \ref{fig:windows} and \ref{fig:vecs} are showing the word2vec model performance. Random Indexing approaches showed a similar pattern when changing window size and vector length, though in the random indexing case, the vector size is set to 2500, larger vectors showed no increase in retrieval performance.

\section{Evaluation}
In order to evaluate the quality of generated timelines, we use the popular ROUGE set of metrics \cite{lin}, which measure the overlap of ngrams, word pairs and sequences between the ground truth timelines, and the automatically generated timelines. ROUGE parameters are selected based on \cite{Owczarzak}. ROUGE-1 and ROUGE-2 are widely reported and were found to have good agreement with manual evaluations in related summarization tasks. In all settings, stemming is performed, and no stopwords are removed. Text is not pre-processed to remove tweet entities such as hashtags or mentions but URLs, photos and other media items are removed. Several ROUGE variants for automatic evaluation are considered, as there is currently no manual evaluation of the generated summaries.

ROUGE-N is defined as the n-gram recall between a ground truth (reference) and system generated summary:
$$\text{ROUGE-N} =$$
$$
\frac
{\sum\limits_{S\in\text{\{Reference Summaries\}}}\sum\limits_{gram_n\in S}Count_{match}(gram_n)}  
{\sum\limits_{S\in\text{\{Reference Summaries\}}}\sum\limits_{gram_n\in S}Count(gram_n)}
$$
where $N$ is the n-gram value, and $Count_{match}(gram_n)$ is the maximum number of n-grams co-occuring in the reference summaries and system generated summary.

ROUGE-L variant is based on the longest common subsequence (LCS) match between the reference and generated summaries. ROUGE-1 and ROUGE-L scores were highly correlated (Pearson's r $> 0.99$), producing the same ranking of system performance.

ROUGE-SU or Skip-bigram variant, measures the overlap of skip-bigrams between summaries. In contrast to LCS, this variant counts all matching word pairs. In the sentence \emph{``Satoshi got free sushi''} has 6 skip-bigrams: [\emph{``Satoshi got'', ``Satoshi free'', ``Satoshi sushi'', ``got free'', ``got sushi'', ``free sushi''}]. Typical settings for the maximum skip distance between two words is set to 4.

A more robust variant of ROUGE, BEwT-E: Basic Elements with Transformations \cite{rougebe} is also reported. Basic Elements are variable sized, syntactically coherent units extracted from text. Transformations are applied to the generated and reference summaries with named entity recognition, abbreviation expansion and others. While this evaluation approach more closely correlates with human judgements in other tasks, the lack of Twitter specific transformations could negatively impact performance - mapping {@}barackobama to ``Barack Obama'' for example. All default BEwT-E settings, part of speech models and named entity recognition models are used.




\subsubsection{Performance on unseen Events}

\begin{table}
\begin{tabu} to \columnwidth {
@{\hspace{0.2em}}X[0.4,l] |
@{\hspace{0.3em}}X[r] |
@{\hspace{0.3em}}X[r] |
@{\hspace{0.3em}}X[r] ||
@{\hspace{0.3em}}X[r] |
@{\hspace{0.3em}}X[r] |[2pt]
@{\hspace{0.3em}}X[r] |
@{\hspace{0.3em}}X[r] |
@{\hspace{0.3em}}X[r] ||
@{\hspace{0.3em}}X[r] |
@{\hspace{0.3em}}X[r] 
}    
    id    & \multicolumn{10}{l}{ROUGE-BEwT Scores} \\ \hline
    ~         & \multicolumn{5}{l|[2pt]@{\hspace{0.3em}}}{Recall} & \multicolumn{5}{l}{Precision} \\ \hline
    
    ~         & \small{BM 25} & \small{w2v dyn.} & \small{RI dyn.} & \small{w2v stat.} & \small{RI stat.} & \small{BM 25} & \small{w2v dyn.} & \small{RI dyn.} & \small{w2v stat.} & \small{RI stat.} \\ \hline

1 & 0.32 & \textbf{0.33} & 0.33 & 0.18 & 0.25 & 0.32 & \textbf{0.33} & 0.33 & 0.18 & 0.25 \\ \hline
2 & \textbf{0.19} & 0.19 & 0.19 & 0.15 & 0.14 & \textbf{0.19} & 0.19 & 0.19 & 0.15 & 0.14 \\ \hline
3 & 0.17 & 0.18 & 0.19 & 0.19 & \textbf{0.20} & 0.17 & 0.18 & 0.19 & 0.19 & \textbf{0.20} \\ \hline
4 & 0.20 & 0.23 & 0.21 & 0.21 & \textbf{0.24} & 0.20 & 0.23 & 0.21 & 0.21 & \textbf{0.24} \\ \hline
5 & 0.14 & 0.17 & 0.16 & \textbf{0.18} & 0.16 & 0.14 & 0.17 & 0.16 & \textbf{0.18} & 0.16 \\ \hline
6 & \textbf{0.19} & 0.17 & 0.16 & 0.15 & 0.14 & \textbf{0.19} & 0.17 & 0.16 & 0.15 & 0.14 \\ \hline
7 & 0.22 & 0.18 & 0.23 & 0.18 & 0.13 & 0.22 & 0.18 & 0.23 & 0.18 & 0.13 \\ \hline
8 & \textbf{0.08} & 0.07 & 0.07 & 0.05 & 0.07 & \textbf{0.08} & 0.07 & 0.07 & 0.05 & 0.07 \\ \hline
9 & 0.20 & 0.19 & 0.21 & 0.14 & 0.16 & 0.20 & 0.19 & 0.21 & 0.14 & 0.16 \\ \hline
10 & \textbf{0.16} & 0.15 & 0.14 & 0.13 & 0.09 & \textbf{0.16} & 0.15 & 0.14 & 0.13 & 0.09 \\ \hline
\end{tabu}
\caption{Detailed Precision \& Recall scores for ROUGE-BEwT for unseen events. Best score in bold.}
\label{fig:evalscores1}
\end{table}

\begin{table}
\begin{tabu} to \columnwidth {
@{\hspace{0.2em}}X[0.4,l] |
@{\hspace{0.3em}}X[r] |
@{\hspace{0.3em}}X[r] |
@{\hspace{0.3em}}X[r] ||
@{\hspace{0.3em}}X[r] |
@{\hspace{0.3em}}X[r] |[2pt]
@{\hspace{0.3em}}X[r] |
@{\hspace{0.3em}}X[r] |
@{\hspace{0.3em}}X[r] ||
@{\hspace{0.3em}}X[r] |
@{\hspace{0.3em}}X[r] 
}    
    id    & \multicolumn{10}{l}{ROUGE-SU4 Scores} \\ \hline
    ~         & \multicolumn{5}{l|[2pt]@{\hspace{0.3em}}}{Recall} & \multicolumn{5}{l}{Precision} \\ \hline
    
    ~         & \small{BM 25} & \small{w2v dyn.} & \small{RI dyn.} & \small{w2v stat.} & \small{RI stat.} & \small{BM 25} & \small{w2v dyn.} & \small{RI dyn.} & \small{w2v stat.} & \small{RI stat.} \\ \hline

1 & 0.15 & 0.14 & \textbf{0.16} & 0.10 & 0.13 & 0.23 & \textbf{0.25} & 0.22 & 0.12 & 0.18 \\ \hline
2 & 0.17 & 0.17 & \textbf{0.19} & 0.17 & 0.16 & 0.40 & \textbf{0.41} & 0.35 & 0.25 & 0.33 \\ \hline
3 & 0.12 & \textbf{0.12} & 0.11 & 0.11 & 0.10 & 0.09 & \textbf{0.10} & 0.09 & 0.09 & 0.09 \\ \hline
4 & 0.14 & 0.17 & 0.15 & 0.16 & 0.17 & 0.20 & 0.25 & 0.23 & 0.25 & 0.27 \\ \hline
5 & 0.12 & \textbf{0.15} & 0.13 & 0.15 & 0.13 & 0.12 & \textbf{0.15} & 0.13 & 0.15 & 0.13 \\ \hline
6 & \textbf{0.22} & 0.21 & 0.19 & 0.20 & 0.16 & 0.19 & 0.20 & 0.22 & 0.20 & 0.23 \\ \hline
7 & \textbf{0.17} & 0.15 & 0.16 & 0.15 & 0.13 & 0.20 & 0.19 & 0.19 & \textbf{0.21} & 0.19 \\ \hline
8 & \textbf{0.08} & 0.07 & 0.08 & 0.05 & 0.07 & 0.20 & 0.31 & 0.22 & \textbf{0.32} & 0.21 \\ \hline
9 & \textbf{0.21} & 0.20 & 0.21 & 0.14 & 0.19 & 0.16 & 0.15 & 0.15 & 0.14 & 0.17 \\ \hline
10 & \textbf{0.12} & 0.10 & 0.11 & 0.10 & 0.07 & 0.30 & \textbf{0.35} & 0.30 & 0.33 & 0.30 \\ \hline

\end{tabu}
\caption{Detailed Precision \& Recall scores for ROUGE-SU4 for unseen events. Best score in bold.}
\label{fig:evalscores2}
\end{table}

%

\begin{figure}
\centering
\small
\definecolor{cC0C0C0}{RGB}{192,192,192}
\definecolor{cC0D0E0}{RGB}{192,208,224}
\definecolor{c1e6823}{RGB}{30,104,35}
\definecolor{c8cc665}{RGB}{140,198,101}
\definecolor{cd6e685}{RGB}{214,230,133}
\definecolor{ceee}{RGB}{238,238,238}
\definecolor{c44a340}{RGB}{68,163,64}
\definecolor{c333333}{RGB}{51,51,51}
\definecolor{c606060}{RGB}{96,96,96}
\begin{tikzpicture}[y=0.50pt,x=0.50pt,yscale=-1, inner sep=0pt, outer sep=0pt]
\input{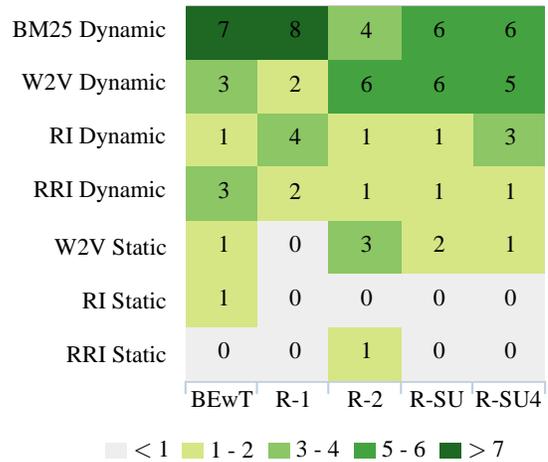}
\end{tikzpicture}
\caption{Methods Ranked 1st in $F_1$ Score, under various ROUGE settings, in all 16 events.}
\label{fig:heatmap1}
\end{figure}

In most cases, shown in Figure~\ref{fig:heatmap1}, adaptive approaches perform well on a variety of events, capturing relevant tweets as the event context changes. This is most notable in the ``WWDC14'' story (Event 10 in Table~\ref{fig:evalscores2}), where there were several significant changes in the timeline as new products were announced for the first time.

While adaptive approaches can follow concept drift in a news story, a notable drawback of DSMs was the lack of disambiguation between multiple meanings of some terms. Even though relevant tweets are retrieved as the news story evolves, irrelevant but semantically related tweets were also present in some timelines - mentions of other car accidents from earlier in the case of the ``Paul Walker'' event for example.

Overall the adaptive NNLM approach performs much more effectively in terms of recall rather than precision. A more effective summarization step could potentially improve accuracy further. This property makes this model suitable for use as a supporting tool in helping journalists find the most relevant tweets for a timeline or liveblog, as the tweets retrieved tend to be much more diverse than those retrieved by the BM25 approach, which favours longer tweets with more repetitive use of terms.

\subsubsection{Diversity of generated Timelines:} 
\begin{figure}
\centering
\small
\definecolor{cC0C0C0}{RGB}{192,192,192}
\definecolor{cC0D0E0}{RGB}{192,208,224}
\definecolor{c1e6823}{RGB}{30,104,35}
\definecolor{cd6e685}{RGB}{214,230,133}
\definecolor{c8cc665}{RGB}{140,198,101}
\definecolor{c333333}{RGB}{51,51,51}
\definecolor{c606060}{RGB}{96,96,96}

\begin{tikzpicture}[y=0.50pt,x=0.50pt,yscale=-1, inner sep=0pt, outer sep=0pt]

\input{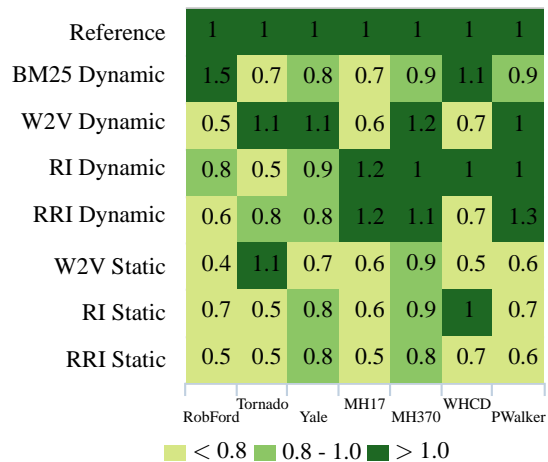}

\end{tikzpicture}

\caption{Diversity relative to reference timelines for events where automatic methods performed better than human generated timelines.}
\label{fig:diversity}
\end{figure}

The average pairwise cosine similarity of tweets in the timelines was used as a measure of diversity. Using the diversity score of the reference timelines, the diversity and redundancy can be compared relative to the available references. Scores above $1.0$ indicate that timelines have less repetition and redundant information than human generated timelines. Scores below $1.0$ indicate that tweets in the timeline are very similar and repetitive. Figure~\ref{fig:diversity} shows events where at least 1 method produces a more diverse timeline than the reference.


The Nonadaptive approach performs well in cases where the story context does not change much, tracking reactions of celebrity deaths for example. Timelines generated with this variant tend to be more general.


While the additive compositionality of learnt word representations works well in most cases, there are limits to this usefulness. Short, focused seed queries tend to yield better results. Longer queries benefit baseline term frequency models but hurt performance of the NNLM approach.

\section{Future and Ongoing Work}
Currently, there is a lack of high quality annotated Twitter timelines available for newsworthy events, as current methods provided by Twitter for creating custom timelines are limited to either manual construction, or through a private API. Other forms of liveblogs and curated collections of tweets are more readily available, but vary in quality. %

As new timelines are curated, we expect that the available set of events to evaluate will grow. We make our dataset of our reference timelines and generated timelines available\footnote{\label{demo}http://mlg.ucd.ie/timelines}.

We adopted an automatic evaluation method for assessing timeline quality. A more qualitative evaluation involving potential users of this set of tools is currently in progress.
%
%
There is also room for improving the model retraining approach. Rather than updating the model training data with a fixed length moving window over a tweet stream, the model could be retrained in response to tweet volume or another indicator, such as the number of ``out of bag'' words, \ie words for which the model does not have vector representations for. Retrieval accuracy is also bound by the quality of our curated tweet stream, expanding this data set would also improve results.


The SumBasic summarization step does not make use of any information from the retrieval models, a better summarization approach that explicitly accounts for diversity and novelty could take better advantage of the DSM approaches.




\section{Conclusion}
Distributional semantic models trained on Twitter data have the ability to capture both the semantic and syntactic similarities in tweet text. Creating vector representations of all terms used in tweets enables us to effectively compare words with account mentions and hashtags, reducing the need to pre-process entities and perform query expansion to maintain high recall. The compositionality of learnt vectors lets us combine terms to arrive at a similarity measure between individual tweets.

Retraining the model using fresh data in a sliding window approach allows us to create an adaptive way of measuring tweet similarity, by generating new representations of terms in tweets and queries at each time window. 

Experiments on real-world events suggest that this approach is effective at filtering relevant tweets for many types of rapidly evolving breaking news stories, offering a useful supporting tool for journalists curating liveblogs and constructing timelines of events.

\section{Acknowledgements}
This publication has emanated from research conducted with the financial support of Science Foundation Ireland (SFI) under Grant Number SFI/12/RC/2289.

We thank Storyful for providing access to data, and early adopters of custom timelines who unknowingly contributed ground truth used in the evaluation.


\begin{quote}
\begin{small}
\bibliography{paper}
\bibliographystyle{aaai/aaai}
\end{small}
\end{quote}

\end{document}